\documentclass[preprint,12pt]{elsarticle}

\usepackage{amssymb}

\usepackage[english,francais]{babel}
\usepackage[latin1]{inputenc}

 \usepackage{amsmath}
 \usepackage{amsfonts}

\journal{Economics Letters}

\begin{document}

\begin{frontmatter}

\title{A longest run test for heteroscedasticity in univariate regression model}




\author[authorlabel1]{Aubin Jean-Baptiste}
\ead{jean-baptiste.aubin@utc.fr}
\author[authorlabel2]{Leoni-Aubin Samuela}
\tnotetext[authorlabel2]{Tel :+ 33 ~ (0)4 ~72 43 72 85}
\ead{samuela.leoni@insa-lyon.fr}

\address[authorlabel1]{Université de Technologie de Compiègne, Rue Personne de Roberval - BP 20529,
  60205 Compiègne, France.}
\address[authorlabel2]{INSA Lyon, ICJ, 20, Rue Albert Einstein,
 69621 Villeurbanne Cedex, France.}
\begin{abstract}
The scope of this paper is the presentation of a test that enables to detect heteroscedasticity in univariate regression model. The test is simple to compute and very general since no hypothesis is made on the regularity of the response function or on the normality of errors. Simulations show that our test fairs well with respect to other less general nonparametric tests.
\end{abstract}


\end{frontmatter}


\section{Introduction}
\label{}
Heteroscedasticity represents
a serious problem in statistics and has to be taken into consideration when performing any
econometric application that could be affected by the latter. Therefore, many statisticians
have put a lot of effort into the elaboration of diagnostic tests enabling accurate detection of
heteroscedasticity.
In particular, a widely and commonly used test for heteroscedasticity is that proposed by
White (1980).
 Note that this test does not presume a particular form of heteroscedasticity.

\bigskip

There are several other tests for the assumption that errors are homoscedastic. If we have prior knowledge that the variance is a linear function of explanatory variables, the Breusch-Pagan test (1979)
 is more powerful. Meanwhile, Koenker (1981)
  proposes a variant of the Breusch-Pagan test that does not assume normally distributed errors. More recently, Luger (2010)
   proposes a simulation-based test with power to detect unspecified forms of heteroscedasticity and two completely nonparametric regression model tests are studied respectively by Dette and Munk (1998)
  and Li et Al. (2006).

\section{Framework}
\label{}
\noindent Consider the sample data $(x_i, Y_i)_{i=1,\hdots,n}$  following the fix-designed regression model:

\begin{equation}
\label{modello}
 Y_i= \mu(x_i)+\sigma(x_i)e_i,~ 1 \leq i \leq n 
\end{equation}

\noindent where $\mu$ and $\sigma$ are unknown functions supported on $\mathcal{S}$. We are interested in testing the problem of homoscedasticity:

$$H_0:~ \exists \sigma_0^2 ~: ~ \forall ~x \in \mathcal{S},~ \sigma^2(x)=\sigma^2_0. $$

\noindent We assume that the errors $(e_i)$ of the model are a sequence of unobserved independent random variables with continuous probability density functions symmetric with respect to $0$ and such that $\mathbb{E}(e_i^2)=1$.\\

Let's consider the 
sequence defined for all $i = 1,\hdots,n$ by:

$$ \varepsilon_i^2 := (Y_i-\widehat{\mu}(x_i))^2, $$
where $\widehat{\mu}$ is an estimate of $\mu$.
\noindent Under $H_0$, $(\varepsilon_i^2)$ is a sequence of realisations of asymptotically independent random variables such that

$$\exists ~ m_0>0 \textrm{ s.t. } \forall~i, ~ \mathbb{P}(\varepsilon_i^2 < m_o^2)= 1/2. $$

\noindent An estimate $\widehat{m}_n^2$ of $m_0^2$ is the median of $(\varepsilon_i^2)_{i = 1,\hdots,n}$. Let's consider the sequence in which the $i$th term is equal to $1$ if $\varepsilon_i^2 \geq \widehat{m}_n $ and $0$ if $\varepsilon_i^2 < \widehat{m}_n $. 
The test statistic is the length of the longest run of 0's or 1's of this sequence, $L_n$. Under $H_0$, if $\left\lfloor n/2 \right\rfloor$ denotes the floor of n/2:

\begin{equation}
\label{legge}
 \forall x =1, \ldots, n, \quad ~\mathbb{P}(L_n \leq x) = S_n^{\left\lfloor n/2 \right\rfloor}(x) / C_n^{\left\lfloor n/2 \right\rfloor},
\end{equation}

\noindent where $S_n^{\left\lfloor n/2 \right\rfloor}(x)$ be the number of sequences of 0's and 1's of length $n$ with $\left\lfloor n/2 \right\rfloor$ 1's such that the length of the longest run of 0's or 1's does not exceed $x$. $C_n^{\left\lfloor n/2 \right\rfloor}$ is the binomial coefficient (and also the number of $n$-sequences containing exactly $\left\lfloor n/2 \right\rfloor$ 1's).\\

\noindent A recursive formula to compute $S_n^{\left\lfloor n/2 \right\rfloor}(x)$ has been presented in Aubin and Leoni-Aubin (prepublication). 
Then, the exact law of $L_n$ can be deduced under $H_0$. Therefore, a value of the test statistic greater than the $(1-\alpha)$th quantile of the deduced distribution indicates rejection of the null hypothesis of homoscedasticity.\\

\section{Results and Discussion}
\label{}
\subsection{Simulation study}
In this section a broad simulation study to illustrate the finite simple performance of the test of homoscedasticity described in the previous section is presented.

To compare our procedure with others in the literature, some of the models simulated in this study are the same as the ones carried by other authors. In particular, the validity and the power of the test proposed in the present paper has been numerically compared with the tests of Dette and Munk (1998)
and of Li et al. (2006)
(denoted respectively in the following DM test and LWI test), choosing some of the regression models used in the simulation study shown in these studies. 
Artificial data are generated according to the  regression model (\ref{modello}),
 with fixed and equally spaced design in the interval [0;1], $x_i=(i-1)/(n-1)$, $i=1, \ldots , n$, indipendent and identically distributed $\mathcal{N}(0,1)$ random errors, and
where $\mu$ and $\sigma$ are given respectively by 

$$\mu(x)=  1+sin(x), \quad ~~ \sigma(x)=  0.5 e^{cx} \qquad ~\qquad~\qquad ~~\textrm{(model 1),} $$
$$ \quad ~ \mu(x)=  1+x,  \quad ~~ \qquad \sigma(x)=  0.5(1+c \sin(10x))^2 \quad ~\textrm{(model 2),} $$
$$~~~\mu(x)=  1+x, \quad ~~\qquad  \sigma(x)=  0.5(1+cx)^2 \qquad ~ \qquad ~~\textrm{(model 3).}  $$

\noindent Table 1 reports the sizes (for different nominal levels) and the powers of the three compared tests for sample sizes $n=50$, $100$ and $c = 0, 0.5, 1$ ($c=0$ is the case in which the homoscedasticity hypothesis is verified). Each table entry is obtained from 1000 independent runs.

\begin{table}[ht]\centering \tiny{
$\begin{array}{ c  c c   c   c c   c    c c   c  c c   }
\hline
& \multicolumn{5}{c}{n = 50} & & \multicolumn{5}{c}{n = 100} \\
\cline{2-6} \cline{8-12} \\
c & \alpha = 4.1 \%  & (\alpha = 5\% )   & & \alpha = 9.8 \% & (\alpha = 10 \% ) & &    \alpha = 5.8 \% & (\alpha = 5\%) &  &  \alpha = 12.5 \% & (\alpha = 10\% ) \\
 \hline 
 \multicolumn{12}{c}{\textrm{Model} 1} \\ 
 \hline \\
 0 & 4.5 & (5 - 5.6)   & & 10.1 & (9.6 - 10.1 ) & &    6.6 & (5.6 - 5.7) &  &  11.8 & (12.6 - 9.3) \\
 0.5 & 7.1 & (11 - 8.4) & &  14.4 & (22.4 - 13.2) & & 10.1 & (12.6 - 9.7) &  & 20.2 & (25.6 - 15.1) \\
 1   &  16.2   & (17.2 - 14.8) & &  28.3   & (28 - 22.3) & &   24.7    & (24.8 - 21.5) & &   37.7   & (45.4 - 31.3) \\
  \hline 
 \multicolumn{12}{c}{\textrm{Model} 2} \\ 
 \hline \\
 0   &   3.9  & (4.8 - 5.3) & &  9.6   & (9 - 10) & &  5.6     & (5 - 4.9) & &  12.7    & (11.6 - 8.9) \\
 0.5   &  24.9   & (32.4 - 27.6) & &  37.3   & (50.4 - 39) & &   41.1    & (55.6 - 43.3) & &   54.7   & (69.6 - 56.8) \\
 1   &  96.4   & (40.2 - 36.5) & &  99.4   & (61.4 - 48.1) & &   100    & (67.6 - 55.7) & &   100   & (79.4 - 67.4) \\
   \hline 
 \multicolumn{12}{c}{\textrm{Model} 3} \\ 
 \hline \\
 0   &   4.6  & (5.2 - 5.4) & &  9.5   & (11.2 - 9.7) & &  6.2     & (5.4 - 5.3) & &  12.6    & (10.6 - 10) \\
 0.5   &  11.2   & (21.4 - 11.3) & &  20.8   & (35.6 -  18.5) & &   18.2    & (17.5 - 15.8) & &   28.6   & (35 - 23.3) \\
 1   &  25.5   & (36.2 - 19.8) & &  40.1   & (54.8 - 29.1) & &   39.4    & (38 - 30.4) & &   55.8   & (53.2 - 41.2) \\
\hline
\end{array}$ }

\label{tavola}
\caption{Empirical sizes and powers of the test $H_0$ with sample sizes 50 and 100.}
\end{table}
\noindent Since the distribution of the test statistic is discrete, we compute the sizes and the powers of the test for nominal levels equal to 4.1\% and 9.8\% in the case $n=50$ (and equal to 5.8\% and 12.5\% in the case $n=100$). For DM and LWI tests, we consider nominal levels equal to 5\% and 10\% for all $n$. So, as far as these tests can be compared, simple inspection of Table 1 allows to deduce that the length of the longest run test is competitive with the DM test. The length of the longest run test fairs well, but LWI test usually gives slightly better powers. Nevertheless, the mean of the obtained powers for the considered cases is higher for the longest run test than for the LWI test. This comes from the fact that in the case 
of the model 2, especially for $n=50$ and $c=1$, the length of the longest run test has a power greatly bigger than the other two tests.\\

\subsection{Conclusion}

\noindent The length of the longest run test gives good results, relaxes several assumptions made in most of other traditional tests and is, in this sense, more general.

\bibliographystyle{elsarticle-harv}







\end{document}